\documentstyle[12pt]{article}
\begin{document}
\title{ORIGIN AND EVOLUTION OF MASSIVE BLACK HOLES IN GALACTIC NUCLEI}
\author{R. D. Blandford\\
130-33 Caltech\\
Pasadena CA 91125 USA} 
\maketitle
\begin{abstract}
Beyond all reasonable doubt, black holes are commonly found in the nuclei
of most normal galaxies.  In recent years, dynamical measurements
of hole masses have transformed the study of their functioning and 
evolution.  In particular, relating their masses, as measured 
contemporaneously,
to the properties of distant quasars can constrain models of
the combined evolution of black holes and their host galaxies. It is suggested
that black hole growth is radiation-dominated 
and demand-limited with an e-folding time of 
$\sim40$~Myr and that most local black hole mass was assembled in AGN
with redshifts, $z>2$, whose counterparts are not directly observed today.
Black hole binaries have additional features and observable 
consequences.
\end{abstract}
\section{INTRODUCTION}
The central problem of developmental biology (for those who are not developmental
biologists) is ``Which came first, the chicken or the egg?''. Likewise a central
question for many of us who do not work regularly on galaxy formation is ``Which came first,
quasars or stars?''. One 
reason why this is important is that a single proton, accreting 
on to a black hole, can 
spawn over a million ultraviolet photons which can ionize 
up to a million hydrogen atoms.
This heating, in turn, controls the scale and the timing of
the collapse of positive density gas fluctuations in the expanding universe and,
ultimately, the formation of the earliest galaxies.

Now the traditional approach to galaxy formation has been largely deductive,
working forward in time from the linear fluctuation spectrum. It is not clear 
how far this approach can be usefully followed in view of the great
complexity of the physical processes involved, and our lack of understanding
of which factors are most important.  By contrast, I believe that it is now more
productive to work, inductively, from direct observational data.  
This introduces several, fresh ingredients, including one that is of great
concern at this meeting, namely that most large galaxies contain surprisingly
massive black holes in their nuclei.
This has direct dynamical implications
for the central stellar distribution. Furthermore, although nuclear activity 
has hitherto been regarded as a sideshow, it now seems distinctly possible
that the formation of a black hole is linked 
to the early evolution of its surrounding galaxy and may 
provide a crucial feedback for limiting star formation, for arresting
its own growth, for moulding the morphology of the galaxy,
for inducing or inhibiting collapse of further density perturbations
in the neighborhood and for 
ionizing the intergalactic medium.  For all of these reasons, it is very important to 
understand the role of black holes in galaxy formation.

The hypothesis that active galactic nuclei (including quasars) are powered by 
accretion onto massive black holes has been around since 1964 and has the immediate
implication that most local galaxies should contain dormant, massive black holes.
(See {\it eg} Krolik 1998 for a discussion of the circumstantial evidence
for black holes in active galactic nuclei that has accumulated over the past
thirty years.) We now know that, beyond all reasonable doubt, local galactic
nuclei contain black holes with masses in the range $\sim10^6$ -
$3\times10^9$~M$_\odot$. Over thirty examples have had their 
masses measured dynamically using a variety
of techniques and with varying degrees of confidence, ({\it eg} Richstone
{\it et al} 1998). This
has transformed the study of AGN as we now can make quantitative the relevant scales 
of length, time, power etc. Three of the most 
precise are NGC~4258 ($3.6\times10^7$~M$_\odot$, Miyoshi {\it et al} 1995), 
our Galactic center ($2.6\times10^6$~M$_\odot$, Genzel \& Eckart 1997)
and M87 ($3\times10^9$~M$_\odot$ Macchetto {\it et al} 1997). 
It appears that most nearby, normal galactic nuclei
contain massive holes. Even more remarkable is the development 
of the capability to measure 
the spin angular frequencies of black holes through the Fe line profiles 
(Tanaka {\it et al} 1995).

Let us now make some convenient definitions. Given a black hole mass, $M$, 
accretion rate $\dot M$ and bolometric luminosity $L$,
we can derive the equivalent length, time and energy
\begin{equation}
m\equiv1.5\times10^{11}M_6{\rm cm}\equiv5M_6{\rm s}
\equiv2\times10^{60}M_6{\rm erg}
\end{equation}
where $M_6=(M/10^6{\rm M}_\odot)$. We can also define the 
fiducial Eddington luminosity, accretion rate and timescale
\begin{eqnarray}
L_{{\rm Edd}}&=&{4\pi GMm_pc\over\sigma_T}\sim3\times10^{10}M_6{\rm L}_\odot\cr
\dot M_{{\rm Edd}}&=&L_{{\rm Edd}}/c^2\sim10^{23}M_6{\rm g\ s}^{-1}\cr
t_{{\rm Edd}}&=&M/\dot M_{{\rm Edd}}\sim0.4{\rm Gyr}
\end{eqnarray}
We also define $\dot m=\dot M/\dot M_{{\rm Edd}}$,
and $\epsilon=L/\dot M c^2$.

For illustration
purposes, I shall adopt a Friedmann universe
with $h\sim0.6,\ \Omega_0\sim0.3,\ \Omega_\Lambda=0$, 
with a total current density $\rho_0
\sim3\times10^{10}$~M$_\odot$ Mpc$^{-3}$, of which $\sim
3\times10^{9}$~M$_\odot$ Mpc$^{-3}$ is baryonic.
The age of the universe at $z=5,3,2,1,0$ is $t\sim1,2,3,7,14$~Gyr,
respectively.
\section{ACCRETION}
It has commonly been supposed that accretion proceeds through a thin disk 
with a radiative efficiency given roughly by the binding energy of the 
smallest, stable circular orbit, $0.06-0.42c^2$, dependent upon
the spin of the hole. In this case, we expect that$\epsilon\sim0.1$ and 
$(L\sim10^{43}\dot mM_6$~erg s$^{-1}$,
where $\dot m\equiv M/\dot M_{{\rm Edd}}$.
However, many observed objects are remarkable for being underluminous relative
to the estimated gas supply.  The case is best made for our Galactic center,
where the supply of gas may be as high as $\sim10^{22}$~g s$^{-1}$
while the bolometric luminosity may be as low as $\sim10^{36}$~erg s$^{-1}$.
The efficiency of conversion of mass into radiant energy may then be 
as small as $\sim10^{-7}c^2$, and is unlikely to be larger than
$\sim10^{-4}c^2$, three to six orders of magnitude below the conventional
value.
Similar claims can be made for other galaxies with measured hole masses.

Observations like these stimulated the development of Advection-Dominated Accretion Flow (ADAF) 
solutions for mass accretion rates well below the Eddington-value,
typically $\dot m<\alpha^2$, where $\alpha$ is the viscosity parameter,
(eg Kato {\it et al} 1998 and references therein). It 
is supposed that there is an efficient viscous torque and that
energy is dissipated into the ions which only heat the electrons through
Coulomb scattering so that the gas flow is essentially adiabatic 
(though not isentropic).
This generally requires that $\dot m<\alpha^2$, where
$\alpha$ is the viscosity parameter. The e-folding time for dormant black holes
is then $>\alpha^{-2}t_{{\rm Edd}}$. 
Models of the spectrum formed by the hot gas in these flows can be adjusted to 
fit the observations of a wide variety of ``underfed'' black hole systems in impressive
detail.  

Despite this success, there are several concerns about the ADAF models. The most fundamental,
dynamical worry is that the flows themselves have positive Bernoulli constant. This 
means that exposed gas has enough energy to escape to infinity.  The fundamental 
reason why this happens is that
viscous torque inevitably transports energy, as well as angular momentum, 
from small
to large radius. An equally important
concern about the radiative model is the assumption that the electrons adopt a
relativistic Maxwellian distribution and avoid non-thermal particle acceleration,
despite the fact that the flows become mildly relativistic,
develop high Mach numbers and rely upon strong, magnetic dissipation to proceed.
A relatively small admixture of suprathermal relativistic electrons 
will greatly increase the emitted flux.

For these, and other, reasons, a generalisation of the ADAF models - the ADiabatic
Inflow-Outflow Solutions (ADIOS) have been developed (Blandford \& Begelman 1998, 1999 in  preparation).
These drop the assumption of conservative mass flow and invoke  a powerful wind
which carries off mass, angular momentum and energy, enabling the remaining
gas to accrete in a bound disk with negative Bernoulli constant. This applies
when the hole is {\it underfed} relative to the Eddington rate.
In the limiting case,
the disk accretion rate at radius $r$ increases, $\dot M\propto r$, 
between the horizon 
of the hole and a transition radius, $r_{{\rm trans}}$, within which the gas is supposed
to be unable to cool. In some cases, {\it eg} the Galactic center, $r_{{\rm trans}}\sim10^5m$
and so it is possible for one proton at $r\sim m$ to sacrifice itself,
altruistically, 
so that $10^5$ of its fellow protons may escape to freedom.  More relevantly, the 
gas density in the vicinity of the hole can be orders of magnitude smaller than in 
the ADAF models and the constraints on the emission correspondingly relaxed. Of course,
real accretion does not have to correspond to this extreme case and rate at which
any outflow carries off mass, energy and angular momentum can be freely assigned.
Not surprisingly, observed spectra can also be fit with ADIOS solutions
(Quataert \& Narayan 1998). We must await
high dynamic range
3D numerical hydromagnetic simulations (Balbus \& Hawley 1998) to make more 
progress. What is important for our purposes is that in ADIOS models there
is a distinction between the rate of mass supply and the rate of mass accretion
and the radiative efficiency with respect to the rate of accretion need not be low.
This means that the black hole mass is quite unlikely to grow during ADIOS accretion.

Similar considerations apply when the accretion is so rapid and the gas flows
so fast that the radiation is trapped. Specifically when $\dot m>1/\epsilon$, the radiation 
is advected out to a radius $r_{{\rm trap}}\sim\dot m m$.  
The flow will be radiation-dominated with $\gamma=4/3$. Again, there are 
ADAF-like solutions ({\it eg} Begelman \& Meier 1982)
where the radiative efficiency is low and black holes can grow rapidly in mass
relative to their radiated energies.  However, these solutions are subject to the 
same dynamical objections as the ADAF solutions and it is likely
that the accretion rate will
again be self-limiting, independent of the rate of mass supply. What this implies, 
in practice, is that when
the black hole is {\it overfed} with gas, it,
will radiate at the Eddington limit and grow with
an e-folding timescale $\epsilon t_E\sim40$~Myr.

There is some observational evidence that this is occuring.
SS433 and the Galactic superluminal sources appear
to have supercritical outflows.  Broad absorption line quasars comprise roughly 
ten percent of radio-quiet quasars and exhibit fast powerful winds that are 
probably accelerated to their terminal velocities by emission line radiation
pressure.  (It is widely believed, though not proven, that all radio-quiet quasars have these flows and
that we only observe them when we lie in the equatorial plane.) However, it is not
yet possible to relate the outflow rates to the hole masses and inflow rates.

For the intermediate, radiative case, $\alpha^2<\dot m<\epsilon^{-1}$, 
there is probably not much mass loss. Traditional, disk accretion ought to be 
appropriate. The hole grows with an 
e-folding time $\sim t_{{\rm Edd}}/\dot m$.
\section{MASS AND ENERGY}
Consider, first, the local universe.  The galaxy 
luminosity density is ${\cal L}_B\sim10^8$
L$_\odot$ Mpc$^{-3}$ ({\it eg} Binney \& Merrifield 1998). 
As the fiducial luminosity $L_B^\ast\sim3\times10^{10}$L$_\odot$,
the bright galaxy density is estimated by $n_G\sim{\cal L}/L_B^\ast
\sim3\times10^{-3}$~Mpc$^{-3}$.
If we adopt a mean stellar mass to light ratio
of $(M/L)_B=6$ in solar units and assume that most
baryons in galaxies are associated with luminous 
stars, then the galaxy baryon mass density 
is $\sim6\times10^8$~M$_\odot$~Mpc$^{-3}$. 
Of this, individual bulge fractions
range from $\sim0.1$ for Scs to unity, for ellipticals,
giving an average bulge mass density of 
$\sim3\times10^8$~M$_\odot$~Mpc$^{-3}$. 
Most of the mass associated with galaxies
is believed to be dark and non-baryonic.  This is very hard 
to measure (and almost as hard to define), but a good guess,
consistent with numerical simulations, puts the mean 
mass density as $\sim10^{10}$~M$_\odot$~Mpc$^{-3}$,
roughly 30 percent of the total mass density.

Turning to the measured black hole masses. ({\it eg}
Richstone {\it et al} 1998). It has been argued that there is a correlation
of the black hole mass, $M_h$ with the bulge mass, estimated photometrically, 
$M_B$, (Magorrian {\it et al} 1998). 
$M_h=0.006M_B$
although the scatter is large.  (Note, also, an intriguing correlation
of $M$ with the radio luminosity reported by Franceschini
{\it et al} 1998.) The local hole mass density is then usually computed from 
the bulge luminosity density and is found to be 
\begin{equation}
\rho_h\sim2\times10^6{\rm M}_\odot {\rm Mpc}^{-3}
\end{equation}
(Note that this estimate is $\propto h^3$.)

We now have a better understanding of the evolution
of bright galaxies.
The earliest galaxies that have been found have $z\sim5.5$ ({\it eg} Spinrad {\it et al} 1998), 
and are observed at an epoch when the universe was $\sim1$~Gyr 
old. It is hard to quantify their density, but they
cannot be too rare, based upon the manner by which they
were discovered.  By the time the universe is $\sim2$~Gyr 
old ($z\sim3$), the density of bright, $\sim L^\ast$, galaxies 
appears to be comparable with their contemporary density, $n_G$,
although uncertainties in the reddening make these estimates
difficult. We do not have good, direct measurements of the 
galaxy luminosity function at  $t\sim3$~Gyr, ($z\sim2$), but can measure a 
small apparent dimming by a factor $\sim3$ in $L_B^\ast$, 
from $t\sim7$~Gyr, ($z\sim1$), to the present, consistent with passive 
stellar evolution following the main star formation epoch
and little change in the number of bright galaxies. However,
there is a marked decrease in the density of low luminosity 
galaxies which outnumber the bright galaxies by a factor $\sim30$ on 
the sky. We do not have a good understanding of the ages and fate
of these faint galaxies, though plausible theories abound.
The star formation rate, which is related to the rate of change of the 
luminosity function, appears to increase
slowly to a value $\sim10^8$M$_\odot$~Gyr$^{-1}$~Mpc$^{-3}$,
when the universe was $\sim4$~Gyr old, ($z\sim1.5$),
and decline by roughly a factor thirty to the present day.

Turning next to quasars, the redshift 5 barrier has, also, been 
broken (Gunn, private communication) and new surveys promise an excellent harvest
with $z>4$. It appears that quasar evolution precedes and 
perhaps exaggerates that of galaxies with a slow increase
in the quasar luminosity to its peak around $t\sim3$~Gyr, ($z\sim2$), 
followed by an {\it apparent} dimming $\propto t^{-3}$.
At the epoch of peak quasar activity, $t\sim3$~Gyr,  
a fraction $\sim0.003$ of bright galaxies is a quasar at any
time $n_Q\sim10^{-5}$~Mpc$^{-3}$. Therefore, if every bright galaxy had a single quasar 
phase then this would last $\sim10$~Myr. 

It has long been recognised that local black hole masses can provide a 
quantitative link to past quasar activity ({\it eg} Lynden-Bell 1969).
Following Soltan (1982), the comoving energy density of emitted quasar light is 
given by
\begin{equation}
U_Q={1\over c}\int dN_Q S(1+z)
\end{equation}
where $S$ is the observed flux and the integral is over all quasars
on the sky.  From the quasar counts, it is apparent that the integral
is just starting to converge at 
$B=22$ (or $\nu S_\nu=5\times10^{-14}$~erg cm$^{-2}$~s$^{-1}$), 
where the number increases at a rate
of just under 2.5 per magnitude. The median redshift of these quasars
is $z\sim2$.  As $N(B<22,z<2.2)
\sim70$ per square degree (Zitelli {\it et al} 1992), we estimate
$U_Q(B)=6\times10^{-18}$~erg cm$^{-3}$ $\equiv100$~M$_\odot$~Mpc$^{-3}$. 
This, of course, only refers to the light observed in the B band and 
emitted at $\sim1500$\AA. For this reason, we must apply a bolometric 
correction to convert it to the bolometric quasar energy density.
This has been variously estimated (on quite insecure grounds)
to lie in the range 10-30. If we choose a value of 20, 
and allow for 50 percent more higher redshift quasars,
we obtain, $U_Q\sim3000$~M$_\odot$ Mpc$^{-3}$, in
agreement with Soltan (1982) and Small \& Blandford (1992),
but a factor 3-6 smaller than obtained by Chokshi 
\& Turner (1992). (The difference appears to be due to 
the inclusion of fainter
quasars that are not directly counted.)  Now, if the black hole 
mass density is built up during the quasar phase by accretion
we can combine Eq. (3), (4) to deduce that the average radiative 
efficiency is $\epsilon\sim0.01$, an order of magnitude smaller than
anticipated, even assuming the Chokshi-Turner value for the radiation
energy density. 
This argument has led many authors to deduce that 
quasars grow while accreting ineffciently.

There is a further way by which we might be underestimating
$U_Q$. Fabian and Iwasawa (1999, and references therein),
have proposed that the X-ray background comprise the combined emission
from many AGN and that its distinctive, hard spectrum, below
$\sim40$~keV may be due to strong absorption.  This hypothesis
implies that much of the $\sim10-100\mu$ infrared background, 
(estimated to be $\sim10^5$~M$_\odot$~Mpc$^{-3}$, Hauser {\it et al} 1998),
be re-radiated ultraviolet and soft X-ray emission produced by accretion
onto black holes, thereby raising $U_Q$ by up to an order
of magnitude.  One concern about this proposal is that most of 
the sources that make up the background must be low redshift AGN rather
than more powerful objects at at high redshift, like the quasars, as
we may have expected.  This is simply because there is unlikely
to be much photoelectric absorption above $\sim30$~keV in the rest frame
of the emitting galaxy, (equivalent to $\sim10$~keV observed for $z\sim2$)
and so, in order to match the break in the background spectrum, it is necessary
that the source redshifts be small. Alternatively, if the break is associated 
with the commonly-observed Compton cut-off energy in Seyfert galaxy 
spectra, which is $\sim100-200$~keV, then redshifts $2<z<4$ are indicated.
(The underlying spectra, below the cut-off energy, would have to be harder than
those associated with local Seyferts, though.)

The Soltan argument may be illusory, because we
probably do not observe directly
the progenitors of the holes that dominate $\rho_h$. To
clarify this point, observe that the bolometric luminosty of
a $B\sim22$, $z\sim2$ quasar is $\sim10^{13}$~L$_\odot$
(continuing to adopt a bolometric correction of $\sim20$).
This is the Eddington limit for a $\sim3\times10^8$~M$_\odot$
hole. If we also adopt an efficiency $\epsilon\sim0.1$, then it takes
$\sim0.1$~Gyr for an accreting quasar to increase its mass by a factor $\sim10$.
suggesting that quasars are active for $\sim 1/30$ of the time at $t\sim3$~Gyr.
Now there are $\sim3\times10^6$ quasars observed,
occupying $\sim3\times10^{11}$~Mpc$^3$
of comoving volume at $z\sim2$. Allowing for 30 times as many inactive
holes, we estimate that the density of $3\times10^8$~M$_\odot$ holes 
at $z\sim2$ is roughly ten percent of the density of bright galaxies.
This is broadly consistent with the observed, local distribution
of black hole mass, specifically, including most of the massive ellipticals
and S0 galaxies. Note that less massive holes, with $M<3\times10^7$~M$_\odot$,
are likely to be too faint to have been detected
as quasars at $z\sim2$. but might have been the unresolved, nuclei of active 
galaxies at $z\sim3$. Indeed most of the growth of lower mass black holes could have 
been rendered invisible by the presence of stellar light. There is
plenty of time to allow black holes to grow from quite small masses 
with e-folding timescales $\sim40$~Myr. 
A more formal treatment, repeating the analysis of Small \& Blandford 1992,
will be presented elsewhere.
\section{FORMATION AND EVOLUTION}
Having suggested an empirical relationship between black holes and their host
galaxies, it is natural to speculate upon how this may have come about.
Under the hierarchical model of galaxy formation, small structures form first
and agglomerate into larger structures. Above the Jeans' mass, bound mass 
perturbations will virialize and shock and then collapse at a rate controlled by the 
rate of cooling and, eventually, the rate of outward transport 
of angular momentum ({\it eg} Haiman \& Loeb 1998).  The 
former is controlled at early times
by the subtle chemistry of molecular hydrogen and at later times
by atomic and ionic line cooling.
The latter is probably dictated initially by gravitational torques associated with departures
from axisymmetry and, ultimately, by magnetic field. Star formation and merging 
with neighboring protogalaxies occurs simultaneously with collapse and the competition
between these various processes will probably only be understood properly from observations.

One scenario (Silk \& Rees 1998, Haehnelt {\it et al} 1998), is that 
a $\sim10^6$~M$_\odot$ black hole forms by coherent collapse in the nucleus 
before most of the bulge gas turns into stars.  The black hole accretes
and radiates at the Eddington limit, driving a wind with kinetic 
luminosity $\sim0.1$ of the radiative luminosity. This deposits energy
into the bulge gas, and will unbind it on a dynamical timescale if 
$0.1L_{{\rm Edd}}>\sigma^5/G$, where $\sigma$ is the bulge velocity 
dispersion. This implies that the black hole mass will be limited to 
a value where it is able to shut off its own fuel supply
\begin{equation}
M<10^5\left({\sigma\over100{\rm km s}^{-1}}\right)^{5/3}{\rm M}_\odot.
\end{equation}
(If it is further assumed that all bulges form dynamically at the 
same time, the $\sigma\propto M_{{\rm bulge}}^{1/3}t^{1/3}$ and
$M\sim M_{{\rm bulge}}^{5/3}$, instead of the linear relation 
proposed above.) 

There are many issues that are unaddressed by this model, including the 
efficiency of star formation, the transport of angular momentum
and radiative cooling.  Furthermore, it is quantitatively inaccurate.
because Eq.~5 does not appear to be satisfied.
({\it eg} $\sigma$(M87)$=330$~km s$^{-1}=2.5\sigma({\rm Galaxy})$,). 
Nonetheless, it does provide a good example of a qualitative 
mechanism whereby the galaxy can limit its own black hole mass.

Indeed, in a radical extension of this idea, a bright AGN may 
also limit infall of gas
to form a disk, though Compton heating, radiation pressure on dust
or direct interaction with a powerful wind.  
When the hole mass and luminosity are large,
the weakly bound, infalling gas will be blown away and an elliptical
galaxy will be left behind.  Only when the hole mass is small,
will a prominent disk develop.  In this case, the bulge to disk ratio, 
(or equivalently the Hubble type), should correlate with the hole 
mass fraction. It would be interesting to perform some numerical hydrodynamical 
simulations that included dynamical energy input asssociated with an AGN.
The recent report by McClure {\it et al} (1999 in press) that essentially
all quasars are associated with elliptical galaxies is in support of
this idea.

The black hole mass may also be limited dynamically. Sellwood \& Moore (1999)
have suggested that strong bars inevitably form in the centers of nascent 
galaxies and channel mass inwards to the growing central black hole
until its mass is $\sim0.02$ of the mass of the disk. At this point,
the bar weakens and infalling mass forms a much more massive bulge which,
in turn, suppresses re-formation of a bar through the creation of an inner 
Lindblad resonance. By contrast, Merritt (1998) has suggested that
the central hole may make the central stellar orbits become 
chaotic, with the consequence that non-axisymmetric 
disturbances are smoothed out and the rate of infall of accreting 
gas falls. These are quite distinct, and no less plausible, mechanisms
by which a black hole can determine galaxy morphology.

In summary, it seems entirely possible that black holes form 
first at quite large redshifts,
$z>>2$ and can grow to their present sizes with standard radiative 
efficiency, by the time of the main quasar epoch at $t\sim3$~Gyr. 
There are several, plausible mechanisms for switching off the 
growth of the hole, some of which have observable signatures.
\section{BLACK HOLE BINARIES}
It is apparent that many high mass galaxies have undergone major mergers 
and, as both partners are likely to 
contain massive black holes, it is almost inevitable that merging black hole
should be quite common.  The 
route to coalescence has been well explored (eg Begelman, Blandford \& Rees
1982, Quinlan \& Hernquist 1997). A captured black hole with mass in excess of $\sim10^6$~M$_\odot$ 
can be dragged into
the galactic nucleus of the larger, 
capturing galaxy by dynamical friction.  Eventually the 
binary will harden when the interior mass is dominated by the host hole and
the orbital speed exceeds the central velocity dispersion.  The evolution
will continue by ejecting low angular momentum stars until the holes
are sufficiently close for the radiation of gravitational waves to dominate the
evolution.

There are several possible and potentially observable implications of these mergers.
Firstly, the dissipative formation of gas-rich galactic nuclei may well account
for the cuspy, power-law central density distributions, with ``disky'' 
isophotes and substantial rotation,  associated with low luminosity
ellipticals and spiral bulges ({\it eg} Faber {\it et al} 1997). If  major merger
subsequently occurs and two massive black holes with their 
stellar entourages merge, in an essentially dissipationless
manner,  this may create the ``cores'' associated with many massive ellipticals.
Secondly, this fits in well with the notion that the giant radio sources
are powered ultimately by black hole spin which may be re-established
in major merger events (Wilson \& Colbert 1995). Thirdly, it is just possible that 
we catch find a rare spectroscopic binary black hole, most plausibly in a low mass,
local Seyfert galaxy ({\it cf} Gaskell 1996).  Fourthly, the actual
coalescences, may be detectable in the future from the gravitational
radiation pulses that they produce through missions like LISA (Bender, these proceedings).
\section{CONCLUSION} 
This is a time of rapid progress in understanding the relationship of black holes
to their hosts. We are starting to piece together a description, initially qualitative,
but now, increasingly quantitative, of how, where and when the massive black holes
were formed. However, the most intriguing question of all is the one with 
which I began. Where were the first ionising photons emitted ({\it eg} Madau
1999)? Was it from 
a population of high mass stars which stimulated cooling and collapse or did galaxies
form from the inside out, growing their nuclear black holes before their stars formed?
We may have to await the next generation of space-borne infrared and submillimeter
telescopes before we are confident of the answer.
\section*{Acknowledgements}
I thank Andy Fabian, Paul Hewett and Richard McMahon for helpful conversations.
The hospitality of the Institute for Advanced Study (through the Sloan
Foundation), the Institute of Astronomy, (through the Beverly 
and Raymond Sackler Foundation)
and NASA, (through contract 5-2837) is gratefully acknowledged.


\begin{thebibliography}{} 
\bibitem{balbus}Balbus, S. A. \& Hawley, J. F. 1998 RMP 70 1
\bibitem{begelman}Begelman, M. C. \& Meier, D. L. 1982 ApJ 253 873
\bibitem{rees}Begelman, M. C., Blandford, R. D. \& Rees, M. J. 1980 Nature 287 307
\bibitem{binney}Binney, J. \& Merrifield, M. 1998 Galactic Astronomy 
        Princeton: Princeton University Press
\bibitem{blandford}Blandford, R. D. \& Begelman, M. C. 1999 MNRAS in press
\bibitem{chokshi}Chokshi, A. \& Turner, E. L. 1992 MNRAS 259 421
\bibitem{fabian}Fabian, A. C. \& Iwasawa, K. 1999 MNRAS 303 L34
\bibitem{faber}Faber, S. M. {\it et al} 1997 AJ 114 1771
\bibitem{franceschini}Franceschini, A., Vercellone, S. \& Fabian, A. C. 1998 MNRAS 297 817
\bibitem{gaskell}Gaskell, C. M. 1996 ApJ 464 L107 
\bibitem{genzel}Genzel, R. \& Eckart, A. 1998 C. R. Acad. Sci. Paris 326 Serie II b 69
\bibitem{hauser}Hauser, M. {\it et al} 1998 ApJ 508 25
\bibitem{haehnelt}Haehnelt, M. {\it et al} 1998 MNRAS 300 817 
\bibitem{haiman}Haiman, Z. \& Loeb, A. 1998 ApJ 503 505
\bibitem{kato}Kato, S., Fukue, J., Mineshige, S., 
          1998, Black Hole Accretion Disks Kyoto: Kyoto University Press
\bibitem{krolik}Krolik, J. H. 1998 Active Galactic Nuclei 
        Princeton: Princeton University Press
\bibitem{lynden}Lynden-Bell, D. 1969 Nature 223 690
\bibitem{macchetto}Macchetto, D. {\it et al} 1997 ApJ 489 579
\bibitem{madau}Madau, P. 1999 Phys. Scripta in press
\bibitem{magorrian}Magorrian, J.{\it et al} 1998 AJ 115 2285
\bibitem{merritt}Merritt, D. 1998 Comm. Ap. 19 1
\bibitem{miyoshi}Miyoshi, M. {\it et al} 1995 Nature 373 127
\bibitem{narayan}Narayan, R., Mahadevan, R. \& Quataert, E. 1998 
\bibitem{quataert}Quataert, E. \& Narayan, R. 1998 astro-ph/9810136
\bibitem{quinlan}Quinlan, G. D. \& Hernquist, L. 1997 New Astronomy 2 533
\bibitem{richstone}Richstone, D. O. 1998 Nature 395 14
\bibitem{sellwood}Sellwood, J. \& Moore, E. M. 1999 ApJ in press 
\bibitem{silk}Silk, J. I. \& Rees, M. J. 1998 A\& A 331 L1
\bibitem{small}Small, T. A. \& Blandford, R. D. 1992 MNRAS 259 725
\bibitem{soltan}Soltan, A. 1982 MNRAS 200 115
\bibitem{spinrad}Spinrad, H. {\it et al} 1998 AJ 116 2617
\bibitem{tanaka}Tanaka, Y. {\it et al} 1995 Nature 375 659
\bibitem{wilson}Wilson, A. \& Colbert, E. J. M. 1995 ApJ 438 62
\bibitem{zitelli}Zitelli, V. {\it et al} 1992 MNRAS 256 349
\end{thebibliography}
\end{document}